\documentclass[aps,prl,twocolumn,superscriptaddress]{revtex4}
\usepackage{mathrsfs}
\usepackage{epsfig}
\usepackage{graphicx}
\usepackage{amsfonts}
\usepackage[figuresright]{rotating}
\usepackage{amssymb}
\usepackage{amsmath}
\usepackage{dcolumn}
\usepackage{bm}
\usepackage{color}

\def\be{\begin{equation}} \def\ee{\end{equation}}
\def\bea{\begin{eqnarray}} \def\eea{\end{eqnarray}}

\newcommand{\WQCASQC} {Wilczek Quantum Center and Key Laboratory of Artificial Structures and Quantum Control, School of Physics and Astronomy, Shanghai Jiao Tong University, Shanghai 200240, China}

\newcommand{\SRCQC}{Shanghai Research Center for Quantum Sciences, Shanghai 201315, China}

\begin{document}
\title{Prethermal time-crystalline spin ice and monopole confinement in a driven  magnet}

\author{Mingxi Yue}
\affiliation{\WQCASQC}

\author{Zi Cai}
\email{zcai@sjtu.edu.cn}
\affiliation{\WQCASQC}
\affiliation{\SRCQC}

\begin{abstract}  
Studies on systems far from equilibrium open up new avenues for investigating exotic phases of matter. A driven-dissipative frustrated spin system is examined in this study, and we suggest an out-of-equilibrium non-magnetic phase where the spins do not order but adhere to the ice rule in space and establish  a long-range crystalline order in time. In contrast to the conventional spin ice, the dynamics of monopoles is confined due to the nonequilibrium feature of our model. Possible experimental realizations of our model has been discussed.

\end{abstract}


\maketitle

{\it Introduction --} Spin ice (SI) is an unusual magnet that does not order even as the temperature tends towards zero\cite{Castelnovo2012}.  Here, geometrical frustration results in ground states with extensive degeneracy yet local constraints known as ice rule. For example, in the rare-earth titanates such as Dy$_2$Ti$_2$O$_7$, the energy is minimized in those configurations satisfying two spins pointing in and two out in each tetrahedra of the pyrochlore lattice\cite{Harris1997,Bramwell2001,Snyder2004}.  Despite its simplicity, the ice rule is responsible for a wealth of intriguing phenomena including the zero-point entropy\cite{Pauling1935,Ramirez1999}, fractionalization\cite{Wan2015,Kourtis2016}, and emergent gauge field\cite{Huse2003,Henley2010}. Locally breaking the ice rule produces a pair of point-like defects -condensed matter analogs of monopoles\cite{Castelnovo2008}- that can be separated to a large distance with a finite energy cost. Most studies on this topic focused on the equilibrium  or near-equilibrium (relaxation\cite{Jaubert2009,Castelnovo2010} or transport\cite{Bramwell2009,Mostame2014,Pan2016}) properties, while the SI physics in far-from-equilibrium systems is elusive. Because the ice rule is rooted in the energy minimization principle, while non-equilibrium systems, especially driven systems, are usually far from  ground states.

Nonequilibrium systems present fresh opportunities for investigating novel phases of matter absent in thermal equilibrium. A prototypical example is the time crystal phase, which spontaneously breaks the temporal translational symmetry\cite{Wilczek2012,Sacha2015,Else2016,Khemani2016,Yao2017,Choi2017,Zhang2017,Mi2022,Frey2022,Kongkhambut2022}. Incorporating spatial degrees of freedom leads to more complex non-equilibrium phases with intriguing space-time structures\cite{Xu2018,Stehouwer2021,Yue2023,Zhao2023}. As for frustrated magnetic systems\cite{Misguich2005}, the role of frustration in a nonequilibrium magnet is still unclear despite great efforts\cite{Wan2017,Wan2018,Bittner2020,Hanai2022,Jin2022}. For example,  one may wonder  whether a magnet driven far from the ground state can host an out-of-equilibrium analog of the SI phase. If so, how does  such a nonequilibrium SI differ from its equilibrium counterpart? Is it possible to define  ``excitations'' above such a non-equilibrium state that has already been highly excited?

In this study, we attempt to answer these questions by investigating a periodically driven classical spin system in a checkerboard lattice. Dynamical simulations of classical spin systems, unlike quantum many-body systems, do not suffer from the notorious exponential wall problem, thus allowing us to simulate 2D systems up to very large system sizes. On the other hand, it has been realized that certain intriguing features of non-equilibrium physics do not crucially depend on the quantum or classical nature of the systems\cite{Yao2020}, and discrete time crystal (DTC) or other exotic orders have been investigated in classical periodically driven systems\cite{Yao2020,Gambetta2019,Khasseh2019,Ye2021,Pizzi2021,Jin2022}. In terms of SI physics, typically, a periodical driving will pump energy into the system thus is detrimental to the SI phase\cite{Slobinsky2010}. Here, we demonstrate that the interplay between  periodic driving and frustration can lead to a non-equilibrium phase that displays oscillating SI patterns in space, accompanied by a DTC order in time. Furthermore, we show that the properties of this non-equilibrium SI phase significantly differs from its equilibrium counterpart since the topological excitation in this case is confined instead of deconfined due to its intrinsic nonequilibrium features.

 \begin{figure}[htb]
\includegraphics[width=0.99\linewidth]{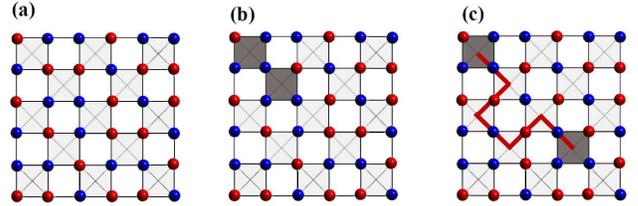}
\caption{(Color online)(a)Schematic  of a perfect spin ice configuration (monopole vacuum) in a checkerboard lattice(blue/red dots indicate spin up/down); (b)Flipping one spin creates two monopole excitations (dark $\boxtimes$) above the vacuum; (c)Two monopoles are separated in space by flipping all the spins in the associated Dirac string (the red line).} \label{fig:fig1}
\end{figure}

\begin{figure}[htb]
\includegraphics[width=0.99\linewidth,bb=1 1 461 178]{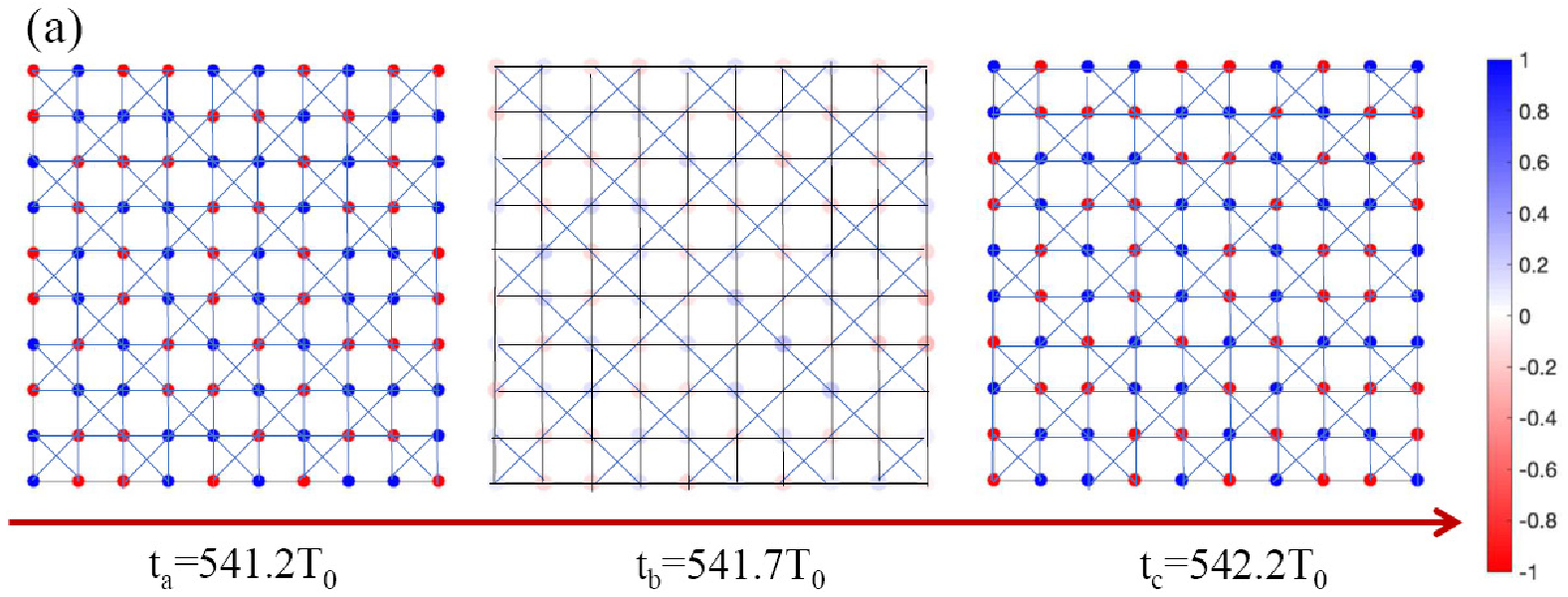}
\includegraphics[width=0.99\linewidth,bb=120 53 1250 580]{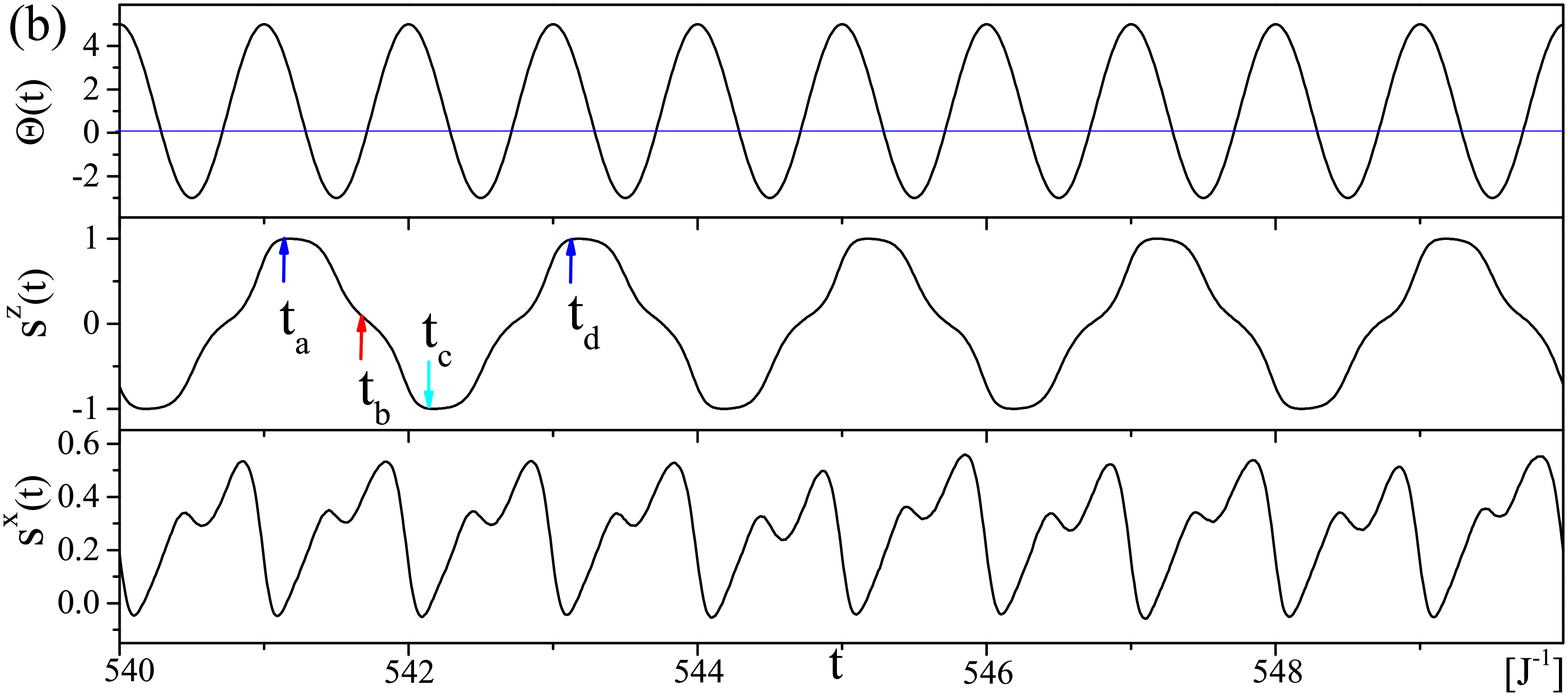}
\includegraphics[width=0.99\linewidth,bb=150 53 1250 580]{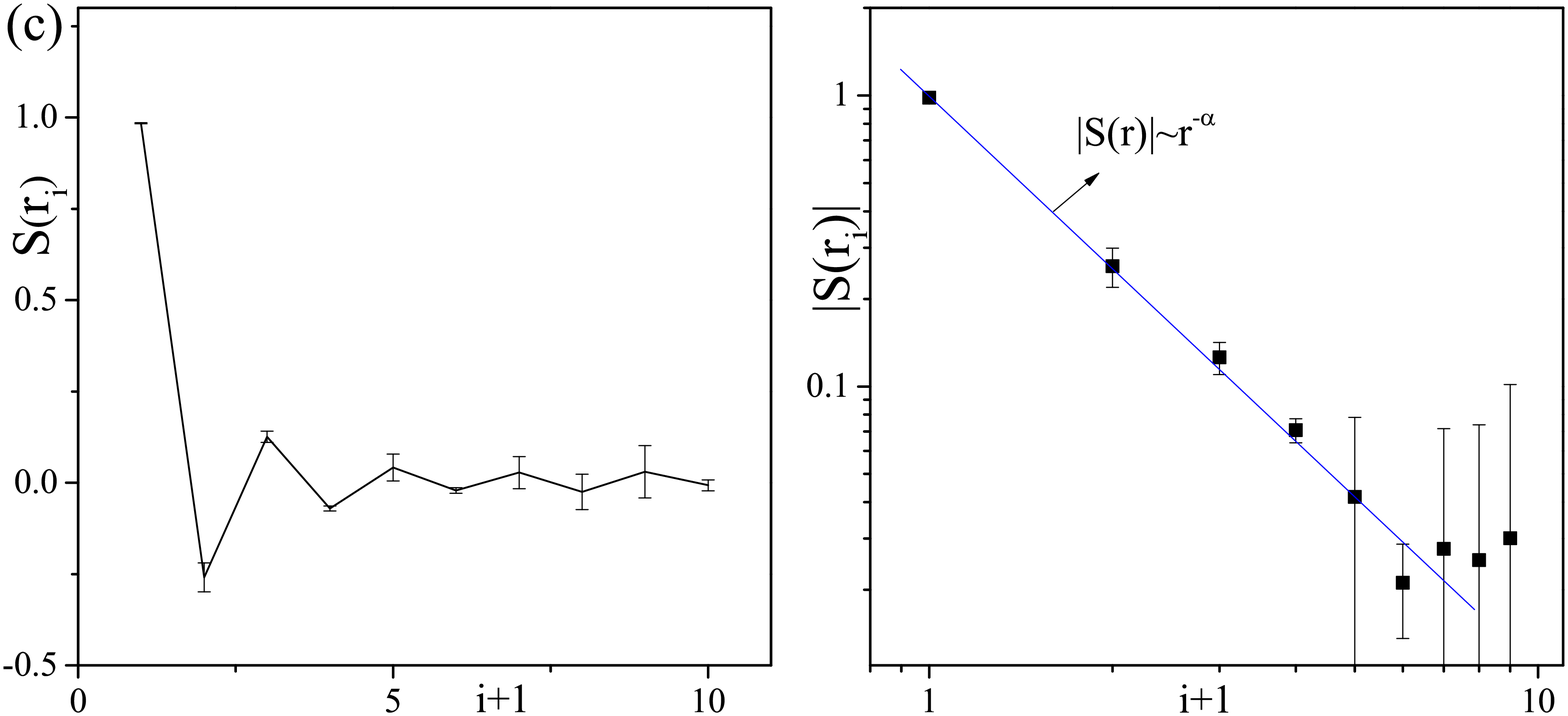}
\caption{(Color online) (a)Snapshots of $\{s_i^z\}$ at three typical time slices; (b)dynamics of the $s_i^z$ and $s_i^x$ of spin on site i; (c)equal-time correlation function at a AF time slice $t=t_a$ (left panel) exhibits an algebraic decay $S(r)\sim r^{-\alpha}$, with the exponent $\alpha=1.9(2)$(right panel). The parameters are chosen as $J'=4J$, $\omega=2\pi J$, $\Gamma=1.5J$, $\gamma=J$, $\mathcal{D}=0.01J$ and $L=30$ (a $10\times 10$ section is plotted here).
} \label{fig:fig2}
\end{figure}

{\it The model --} To examine the spin ice phase, we employ a classical transverse Ising model in a checkerboard lattice, whose Hamiltonian reads:
\begin{equation}
H_{ice}=\sum_{\boxtimes}\sum_{ij\in \boxtimes } \Theta (t)s_i^z s_j^z+\Gamma\sum_i s_i^x, \label{eq:Ham}
\end{equation}
where $\boxtimes$ indicates the plaquette in the checkerboard lattice with the next nearest neighboring (NNN) coupling (the grey plaquette in Fig.\ref{fig:fig1} a). $\mathbf{s}_i=[s_i^x,s_i^y,s_i^z]$ is a  classical vector with a fixed length $|\mathbf{s}_i|=1$. $\Gamma$ is the strength of a time-independent transverse field, which is important not only for the realization nontrivial spin dynamics (without $\Gamma$, the spin dynamics is simply a precession around z-direction), but also for the confinement feature of the monopole dynamics, as we will show in the following.  $\Theta(t)=J+J'\cos \omega t$ is a periodically varying interaction strength, where $\Theta(t)$ being positive/negative indicates anti-ferromagnetic(AF)/ferromagnetic(FM) coupling. $J'$ and $\omega$ represent the amplitude and frequency of the driving. Here, we fix these Hamiltonian parameters. However, we shall demonstrate in the supplementary material(SM)that the key results of this work do not crucially depend on this specific choice of parameters\cite{Supplementary}.

 Typically, periodic driving will heat  closed interacting systems towards an infinite temperature state. To avoid this avoid this featureless asymptotic state, we introduce dissipation into our model by coupling each spin to a thermal bath, which can be phenomenologically modeled via stochastic methods. In the presence of a thermal bath, the dynamics of spin $i$ can be described by a stochastic Landau-Lifshitz-Gilbert equation\cite{Brown1963}:
\begin{equation}
\dot{\mathbf{s}}_i=\mathbf{h}_i(t)\times \mathbf{s}_i- \gamma  \mathbf{s}_i\times (\mathbf{s}_i\times \mathbf{h}_i(t)) \label{eq:EOM}
\end{equation}
where $\gamma$ is the dissipation strength, which is fixed as $\gamma=J$ for the numerical convenience. Although this value is larger than that in conventional magnet, the long-time asymptotic state does not importantly depends on  $\gamma$\cite{Supplementary}.  $\mathbf{h}_i(t)=\mathbf{h}^0_i(t)+\bm{\xi}_i(t)$, where $\mathbf{h}^0_i=-\nabla_{\mathbf{s}_i} H_{ice}=[\Gamma,0,-\Theta(t) \langle s_i^z\rangle]$  is the effective magnetic field on site $i$ and $\langle s_i^z\rangle=\sum_j s_j^z$ where the summation is over all the six neighboring spins of site i. $\bm{\xi}_i(t)$ is a 3D zero-mean random field  representing thermal fluctuations.  The local bath  satisfies: $\langle \xi_i^\alpha(t)\xi_j^\beta(t')\rangle_{\bm\xi}=\mathcal{D}^2\delta_{\alpha\beta}\delta_{ij} \delta(t-t')$
where $\alpha,\beta=x,y,z$  and $\mathcal{D}$ is the strength of the noise. If the bath is in thermal equilibrium, $\gamma$ and $\mathcal{D}$ should satisfy  $\mathcal{D}^2=2T \gamma$, where $T$ is the temperature of the bath. The stochastic differential equation can be numerically  solved by  the standard Heun method with a Stratonovich's discretization formula\cite{Ament2016}, in which we select the discrete time step $\Delta t=10^{-3}J^{-1}$ (the convergence with smaller $\Delta t$ has been verified). The simulation is performed over a $L\times L$ checkerboard lattice with periodic boundary condition, and we choose  random initial states  whose effect has also been analyzed\cite{Supplementary}. In the following, we will focus on the long-time asymptotic dynamics of this model, whose  dynamical phase diagram  is extremely rich\cite{Supplementary}. Here, we only consider the scenario when the system concurrently displays SI patterns in space and DTC order in time, as opposed to listing all the dynamical phases.

{\it Time-crystalline spin ice --} We consider the case where $\Theta(t)$ oscillates between the AF and FM couplings (this condition, however, is not necessary for the TC-SI phase, which can exist even when $\Theta(t)$ does not change its sign, see the SM\cite{Supplementary}), and the spin configuration accordingly varies. The snapshots of $\{ s_i^z\}$ at three typical time slices have been plotted in Fig.\ref{fig:fig2} (a).  At a time slice $t_a=541.2T_0$ with AF coupling  ($T_0=1/J$ is the period of $\Theta(t)$, the $s_i^z(t)$ has a $0.2T_0$ phase lag with respect to $\Theta(t)$), each $s_i^z$ reaches its maximum ($|s_i^z|=0.9994$),  and $\{ s_i^z\}$ obeys the ice rule ($\sum_{ij\in \boxtimes} s_i^z$ vanishes for all $\boxtimes$). The  $\{ s_i^z\}$ at the following FM time slice $t_b=t_a+0.5T_0$ ($\Theta(t_2)<0$) shows neither spin ice pattern, nor FM  order, rather, it is a paramagnetic phase (PM) with magnetization along the x-direction (see Fig.\ref{fig:fig2} b). At the time slices $t_c=t_a+T_0$, the system Hamiltonian return to the original one $(H_{ice}(t_c)=H_{ice}(t_a))$, but $\{ s_i^z\}$ does not. Instead, all of them are  simultaneously reversed $\{s_i^z(t_c)\}=\{-s_i^z(t_a)\}$, thus the ice rule is still preserved. $\{ s_i^z\}$ return to its original values after two periods of driving at $t_d=t_a+2T_0$ ($\{s_i^z(t_d)=s_i^z(t_a)\}$), which indicates a spontaneous $\mathbb{Z}_2$ time translational symmetry breaking (TTSB).

The origin of the DTC order can be understood as a consequence of the periodically driven interaction. For a pair of adjacent sites $ij$, if $s_i^z(t)$ and $s_j^z(t)$ synchronize as $s_i^z=s_j^z \sim \cos[\omega't+\phi]$,  the instantaneous interacting energy  $H_I(t)\sim \Theta(t)\cos[2\omega't+2\phi]$ with  $\Theta(t)\sim \cos\omega t$ can be expressed as    $H_I(t)\sim \cos[\delta\omega t-2\phi]+\cos[\Omega t+2\phi]$ with $\delta\omega=\omega-2\omega'$ and $\Omega=\omega+2\omega'$. $H_I(t)$ oscillates around zero except for the period doubling case ($\omega'=\omega/2$), where $H(t)\sim \cos2\phi$ (the fast oscillating term $\cos[2\omega t+2\phi]$ is omitted).  Therefore $H_I(t)$ becomes approximately time-independent and takes its minimum value at two degenerate points $\phi_1=\frac \pi2$ and $\phi_2=\frac{3\pi}2$,  which is  responsible for the spontaneous $\mathbb{Z}_2$ TTSB in the DTC. This intuitive picture also explains the fact  only $\{s_i^z\}$  exhibit period doubling, while $\{s_i^x\}$ do not (because the periodic driving is imposed on the interaction $\Theta(T)s_i^z s_j^z$ instead of on the external field $\Gamma s_i^x$), as shown in Fig.\ref{fig:fig2} (b).

The equilibrium SI supports a Coulomb phase with an algebraic decay of the spatial correlation function, one may query  whether this property holds for the TC-SI phase. To answer this question, we calculate the equal-time correlation function $S(\mathbf{r})=\frac 1{L^2}\sum_\mathbf{i}\langle s_\mathbf{i}^z(t_a)s_{\mathbf{i}+\mathbf{r}}^z(t_a)\rangle$ at an AF time slices ($t=t_1$). Fig.\ref{fig:fig2} (c) indicates that along the diagonal direction $\mathbf{r}=\frac 1{\sqrt{2}}(r,r)$ with $r=|\mathbf{r}|$, $S(r)$ decays algebraically in distance $S(r)\sim r^{-\alpha}$, with $\alpha=1.9(2)$ agreeing with the exponent predicted by the Coulomb phase\cite{Henley2010} ($\alpha=d$ with $d=2$ the dimension of the lattice). However, this agreement does not indicate that the asymptotic state in our model adiabatically follows the ground state of the $H_{ice}$. First, the ice rule only hold at the time slices with AF coupling. For example, at a time slice with FM coupling ({\it e.g.} $t=t_b$), the ground state of $H_{ice}(t_b)$ is supposed to be an FM state along the z-direction, while our system actually exhibits a PM state. Furthermore, the spontaneous TTSB is forbidden in thermal equilibrium due to the no-go theorem\cite{Bruno2013,Watanabe2015}. Therefore, the asymptotic state in our model is a genuine non-equilibrium state with alternating SI  and PM configurations in space and DTC order in time.

 \begin{figure}[htb]
\includegraphics[width=0.99\linewidth]{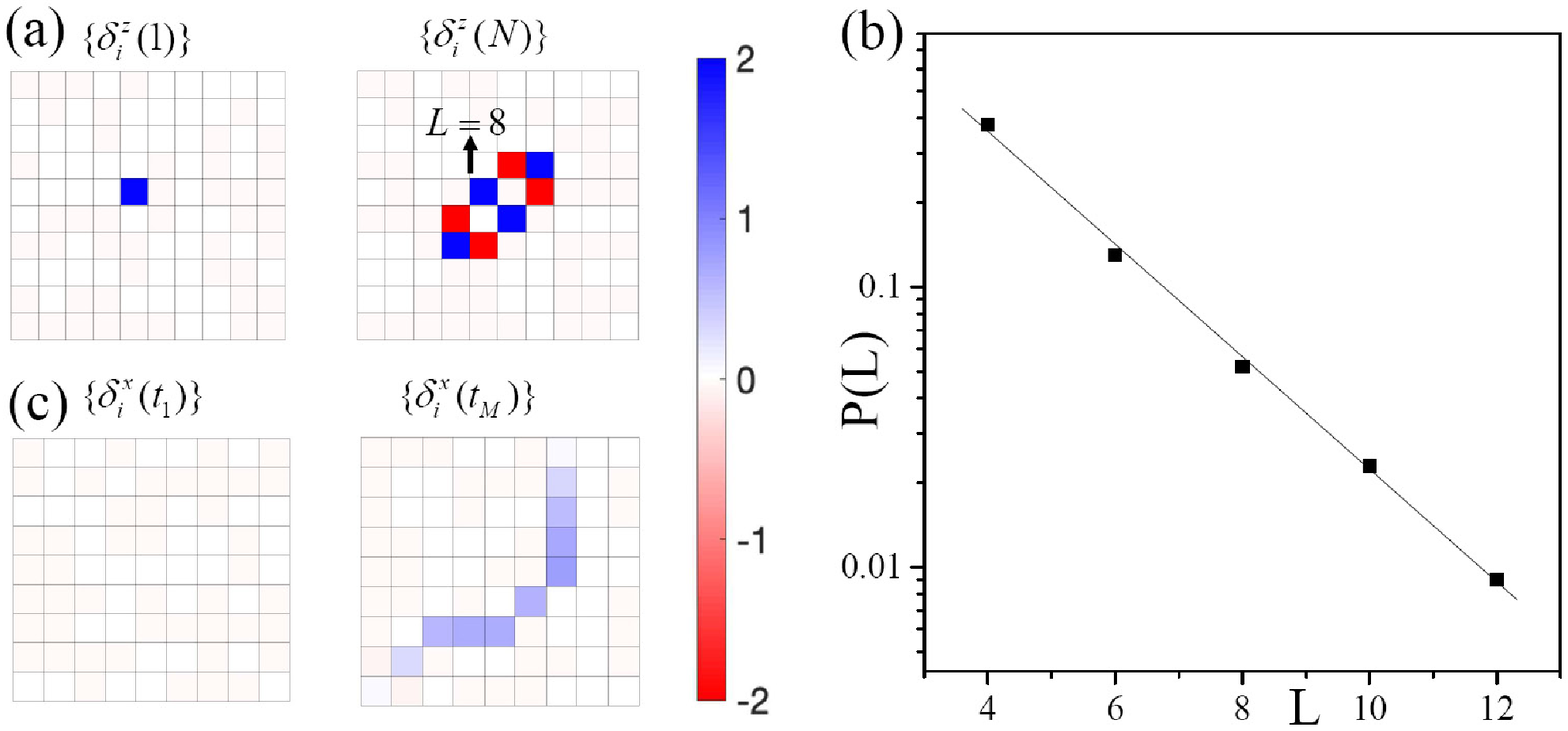}
\includegraphics[width=0.99\linewidth,bb=184 58 1170 580]{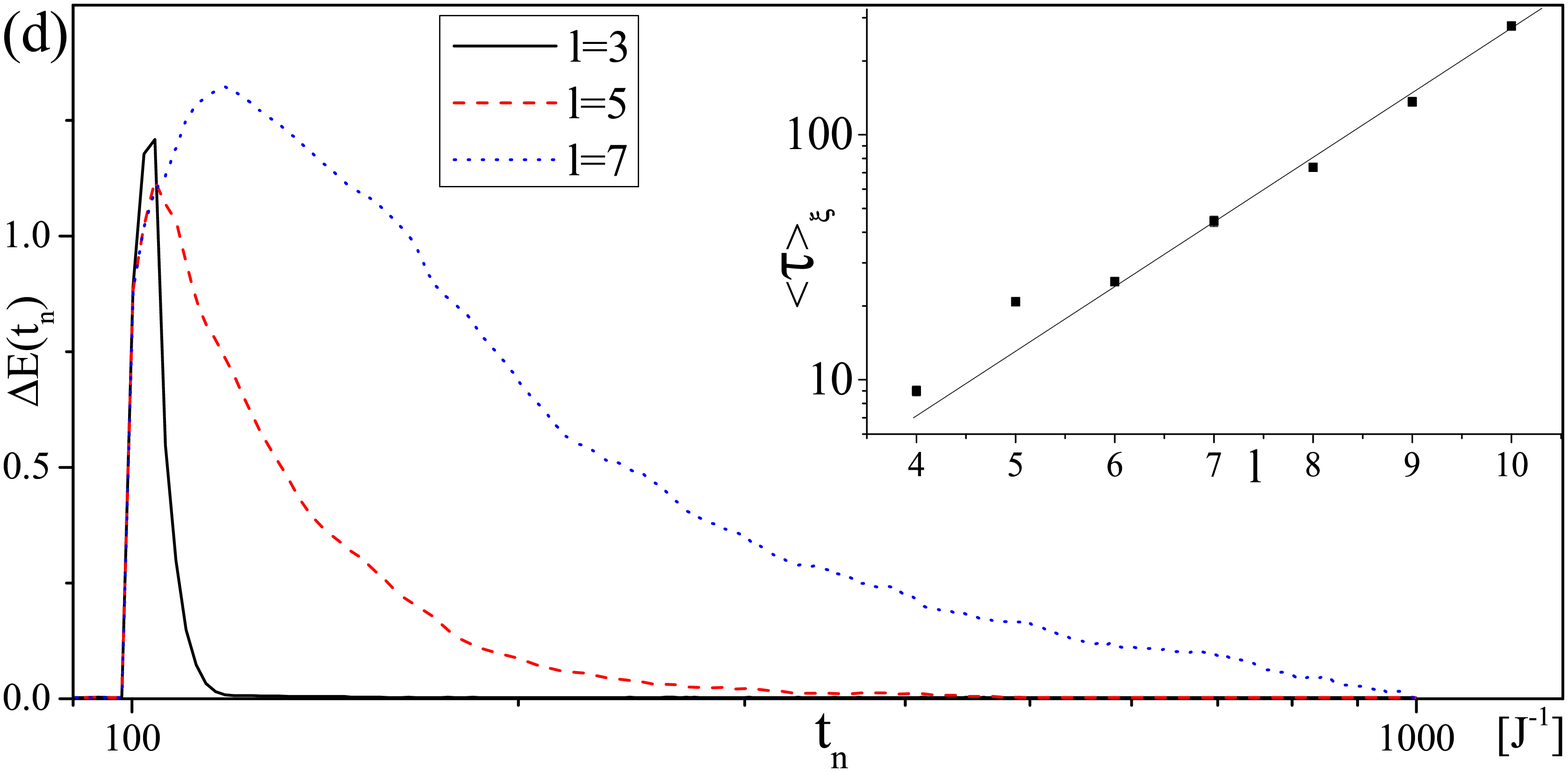}
\caption{(Color online) (a) The difference between  $s^z$-configurations with and without a single spin flip $\{\delta_i^z(t)\}$ at the time $t=t_1$ (initial state) and $t=t_N$  (final states). (b)The probability distribution in terms of  the circle length $L$ in the final states.  (c) The  difference between the $s^x$-configurations with and without a single spin flip $\{\delta_i^x(t)\}$ at the time $t=t_1$  and $t=t_M$ (intermediate state). (d)The energy difference $\Delta E(t_n)$ at the AF time slices $t_n=t_0+2T_0 (n-1)$ starting from different initial states, each of which contains one Dirac string with different length $l$. The inset indicates the average relaxation time $\langle \tau\rangle_{\bm\xi}$ as a function of the length of the Dirac string in the initial state. Other parameters are chosen the same as in Fig.\ref{fig:fig2}. (a)-(c) starts from an initial state with a single $s_z$ flip, and (d) from states with a Dirac string.} \label{fig:fig3}
\end{figure}

{\it Confined dynamics of monopoles--} In a conventional SI, the elementary excitations can be introduced by flipping one spin in a perfect SI configuration, which violates the ice rule in the two adjacent $\boxtimes$. For a monopole ``vacuum'' (a perfect SI configuration), flipping a spin equals to create of a pair of monopoles, which can be separated by properly flipping a chain of spins (the Dirac string\cite{Morris2009} as shown in Fig.\ref{fig:fig1} c).  The energy required to separate two monopoles in a conventional SI model with short-range coupling is independent of their distance, therefore the monopoles herein are deconfined. Next, we will demonstrate how the nonequilibrium features of our model qualitatively alter this deconfined scenario of  monopoles.

Even though the definition of ``excitation'' above an out-of-equilibrium state is elusive, for our model, we  adopt a similar procedure of exciting the state by flipping one spin, and monitor its subsequential dynamics. For this purpose, we first choose an AF time slice $t_1$ when all $s_i^z$  reach their maximum and the corresponding $s^z$ configuration obeys the ice rule. Then we randomly pick a site (say, site $i$) and flip its  z-component to obtain a configuration  $\{s^z_j(1)\}$, then study its subsequential dynamics $\{s_j^z(t)\}$  and compare it to the dynamics without spin flip $\{\bar{s}_j^z(t)\}$. At the AF time slices $t_n=t_1+2T_0 (n-1)$, we defined $\delta_j^z(n)=s_j^z(t_n)-\bar{s}_j^z(t_n)$ to measure the $s^z$ difference between the $s^z$ configurations with and without spin flip. At $t=t_1$, only $s_i^z$ is flipped, and thus $\delta_j^z(1)=0$ except $j=i$. Due to dissipation,  after sufficiently long time ($t_n>t_N$ with $t_N$ being the relaxation time), the system will reach a new SI configuration. The difference configuration $\{ \delta_j^z(N)\}$ plotted  in Fig.\ref{fig:fig3} (a) shows that the flipped spins during this process form a circle with alternating + and $-$ structure. Physically, flipping one spin produces two monopoles, each of which can propagate from one $\boxtimes$ to another by flipping the spin between them. A monopole is a topological fractionalized object that can not be created or  annihilated by itself. As an alternative, monopoles can only be annihilated in pairs when they collide during the propagation, which will leave behind a new SI configuration that differs from the original one by flipping all of the spins along the trajectories the two monopoles went through.

The monopole dynamics in the TC-SI phase seems similar to a random walk in conventional SI, where the monopoles are deconfined. However, we will show that it is not the case. To this end, we study the dynamics after a single spin flip under different noise trajectories, and calculate the statistical distribution $P(L)$ of the circle length $L$ in the final states. As shown in Fig.\ref{fig:fig3} (b),  $P(L)$ decays exponentially with L, which implies a localized feature of the monopole trajectories. On the contrary, in the conventional SI, the monopoles are delocalized, thus their dynamics are similar to a random walk, which suggests  an algebraic decay of $P(L)$.

The striking difference between the TC-SI and conventional SI is due to the non-equilibrium feature of our system, which does not always in a SI phase, but keeps oscillating between the SI and PM phases during evolution. Even though separating two monopoles does not cost energy in the SI phase, it indeed requires energy proportional to the distance between the monopoles in the PM phase. To verify this point, we calculate the difference between the  $s_x$-configurations with ($\{s_j^x(t)\}$) and without ($\{\bar{s}_j^x(t)\}$) the single spin flip:
$\delta_j^x(t)=s_j^x(t)-\bar{s}_j^x(t)$. As shown in Fig.\ref{fig:fig3} (c), initially, since only the $s_i^z$ is flipped, the $s_x$-configurations are the same ($\delta_j^x(t_1)=0$).  We then choose an intermediate time slice $t=t_M$, when the interaction energy happens to vanish ($\Theta(t_M)=0$) and the two monopoles have not annihilated. As shown in Fig.\ref{fig:fig3} (c), when two monopoles are separated in distance, the motion of the monopoles will leave a string with nonvanishing $\delta_j^x(t_M)$ along the trajectories they went through, which will cost an energy proportional to the length of this string ($\Delta E(t_M)=\sum_{j\in string} \Gamma \delta_j^x(t_M)$), resulting in the monopole confinement.

 One can also start the dynamics from an initial state with a pair of well-separated monopoles attached by a Dirac string (at $t=t_1$, we flip a string of spins  as shown in Fig.\ref{fig:fig1} (c), and study the subsequential dynamics).  We monitor the excess energy $\Delta E_n=\langle E(t_n)-\bar{E}(t_n)\rangle_{\bm\xi}$, where $\langle \rangle_{\bm\xi}$ is the ensemble average over the thermal noise trajectories, and $E(t_n)$ and $\bar{E}(t_n)$ indicate the instantaneous energy with and without the string spin flip.  Fig.\ref{fig:fig3} (d) shows that the relaxation is slower from an initial state with a pair of monopoles with larger separation, and its inset indicates that the average relaxation time $\langle \tau\rangle_{\bm\xi}$ exponentially diverges with the length of the Dirac string $l$. This exponentially long life-time of a Dirac string agrees with the confinement-induced localization of the monopole dynamics as analyzed above.

\begin{figure}[htb]
\includegraphics[width=0.99\linewidth,bb=150 53 1350 580]{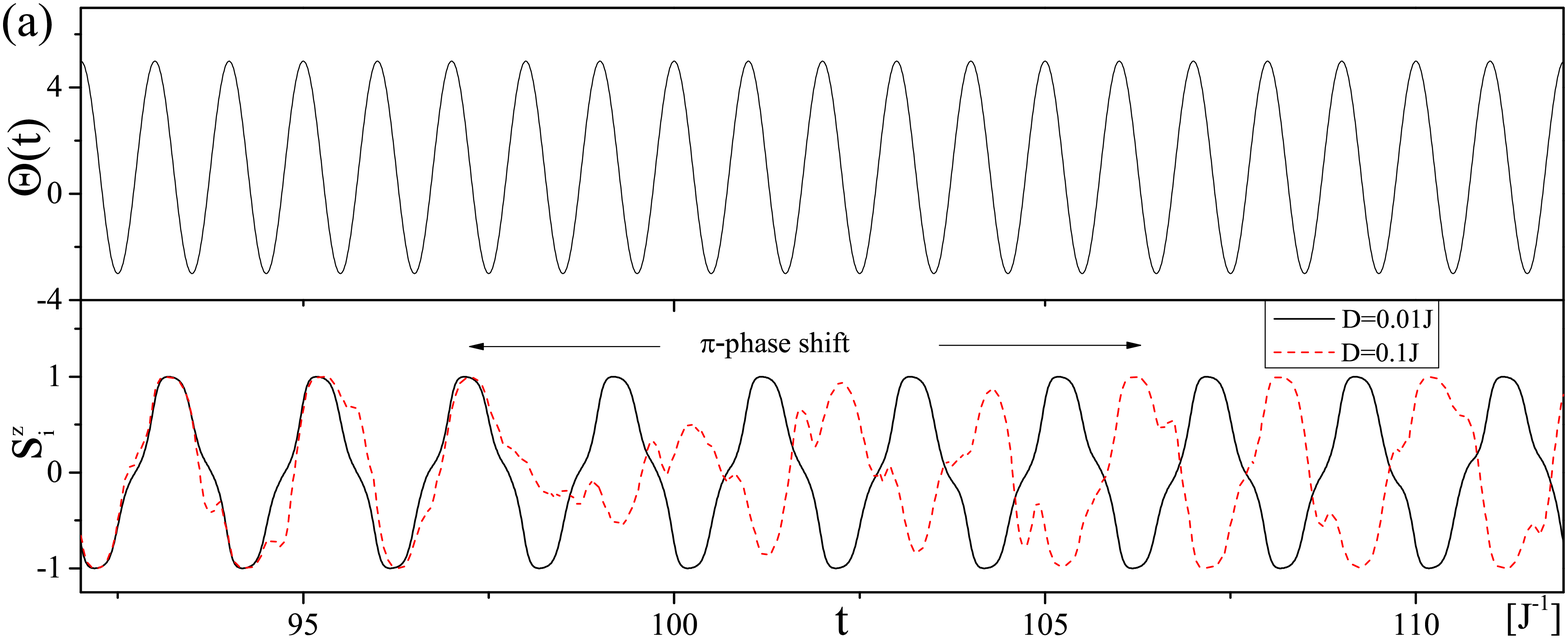}
\includegraphics[width=0.99\linewidth,bb=190 90 1300 580]{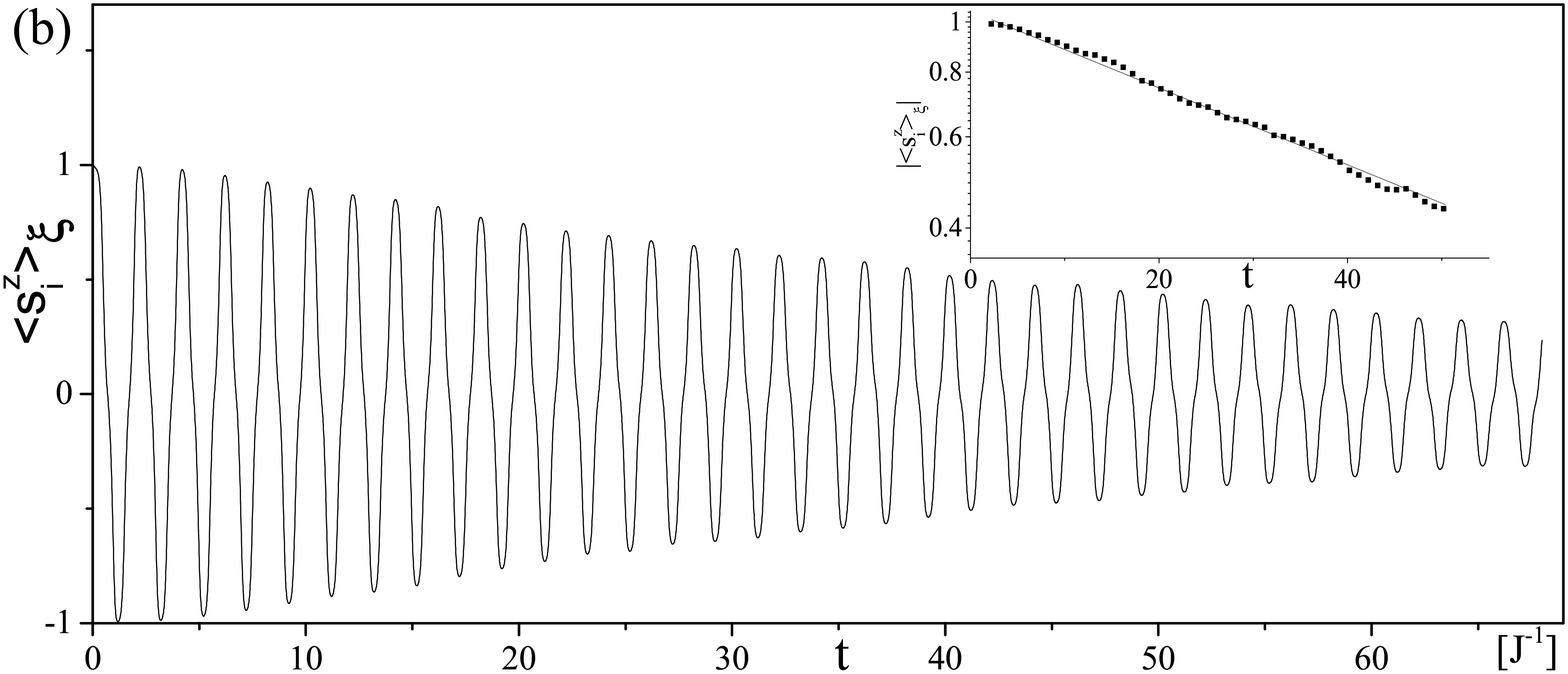}
\caption{(Color online) (a) The dynamics of $s_i^z$ on site $i$ in a single noise trajectory with $\mathcal{D}=0.01J$ and $\mathcal{D}=0.1J$, the former demonstrates a perfect DTC order, while in the latter, thermal fluctuations activate a $\pi$-phase shift; (b) The dynamics of the average $\langle s_i^z\rangle_{\bm\xi}$ starting from a perfect SI state after ensemble average over $10^3$ noise trajectories. The envelope of $\langle s_i^z\rangle_{\bm\xi}$ exhibits an exponential decay as shown in the inset. Other parameters except $\mathcal{D}$ are chosen the same as in Fig.\ref{fig:fig2}.
} \label{fig:fig4}
\end{figure}

{\it Instanton activated by the thermal fluctuation --} Besides the confinement of the monopole dynamics, the TC-SI phase is also distinct from its equilibrium counterpart due to the presence of DTC order. One may wonder how the monopoles affect the DTC order of the TC-SI phase. The answer to this question is related to the stability of TC-SI phase against thermal fluctuations, which excite monopoles with a finite density. The Coulomb phase in equilibrium SI does not breaks any symmetry, and is not robust at finite temperature. However, the TC-SI phase exhibits a spontaneous $\mathbb{Z}_2$ TTSB, while a discrete symmetry breaking phase is typically assumed to be robust against weak thermal fluctuations in 2D systems. For example, in a similar but non-frustrated model, the corresponding DTC phase is indeed stable at low temperature\cite{Yue2022}. The impact of thermal fluctuation on the TC-SI phase will then be discussed.

 Unlike the conventional SI phase,  once a spin in our TC-SI phase is suddenly flipped at a typical AF time slice, it does not only produce a pair of monopoles in space, but also results in a $\pi-$phase shift on top of the periodic dynamics of this flipped spin, which corresponds to  tunneling from one ``degenerate'' DTC phase ($\phi=\frac \pi2$) to the other ($\phi=\frac{3\pi}2$). Such a fluctuation-activated tunneling  between the two  $\mathbb{Z}_2$ symmetry breaking states (see Fig.\ref{fig:fig4} a) resembles the instanton excitation in the field theory\cite{Rajaraman1987}, and is a topological defect in the temporal domain. These instanton excitations, no matter how rare they are,  are detrimental to the DTC long-range order and result in an exponential decay of $s_i^z$,as shown in Fig.\ref{fig:fig4} (b). Therefore, at any finite temperature, the proposed TC-SI is actually a prethermalized phase, however, its life-time can be extraordinarily long at a temperatures much lower than the activated temperature of monopoles.

{\it Experimental realizations of dynamically modulated interactions--} One of the primary obstacles for experimental realization of our model is that it requires a periodical driving imposed on the interaction rather than on the external field, which seems unrealistic for solid-state magnets.  However, such a dynamically modulated magnetic interaction can be achieved using  magnetophononics, in which the electric field of a laser is coupled to the phonon, and the consequent periodic atomic displacements could dynamically modulate the magnetic exchange couplings between the spins\cite{Yarmohammadi2021,Deltenre2021}. This proposal has been realized experimentally in the AF semiconductor $\alpha$-MnTe\cite{Bossini2021}.  Although the tunable coupling regime therein  is small and it is impossible to change the sign of the interaction, we show that, for a slower driving ({\it e.g.} $\omega=0.5\pi J$) the TC-SI can exist even when the coupling is always AF ($J'<J$)\cite{Supplementary}.  This periodically modulated interaction is also  accessible in synthetic quantum systems such as trapped ions\cite{Schneider2012} and cavity QED systems\cite{Gopalakrishnan2009}, where  the magnetic interaction mediated by cavity can be dynamically controlled by applying a periodic driving to the cavity photons. Introducing quantum fluctuation might give rise to interesting order by disorder phenomena in equilibrium\cite{Shannon2004,Shannon2012,Henry2014}, and a quantum generalization of our non-equilibrium model might  provide new perspectives for the SI physics.


{\it Conclusion and outlook --} In summary,  by studying a driven-frustrated magnet, we reveal an intriguing confinement mechanism, which is rooted in the non-equilibrium feature of the system.  It is shown that for a non-equilibrium system driven far from the adiabatic limit, the instantaneous states at different time slices ({\it e.g.} the SI and PM states in our case) might not be independent with each other, instead, they can build up temporal correlations and affect each other. For instance, the PM states in our model effectively mediate the monopole interactions in the SI states,   and make them significantly differ from their equilibrium counterparts.   This  general picture behind the proposed confinement mechanism is not restricted to the specific model in this study, but can apply to a wide class of non-equilibrium classical and quantum systems which exhibit alternating phases during the time evolution. As for the frustrated quantum magnetism, a similar phase without spontaneous symmetry breaking is the quantum spin liquid. One thus may wonder whether it is possible to realize similar exotic quantum phases of matter out of equilibrium\cite{Jin2023}, which can simultaneously show spatial topological order and non-trivial temporal (long-range or quasi-long-range) orders.

{\it Acknowledgments}.--- This work is supported by the National Key Research and Development Program of China (Grant No. 2020YFA0309000), NSFC of  China (Grant No.12174251), Natural Science Foundation of Shanghai (Grant No.22ZR142830),  Shanghai Municipal Science and Technology Major Project (Grant No.2019SHZDZX01).


\end{document}